\documentclass[%
superscriptaddress,
preprint,
showpacs,
nofootinbib,
nobibnotes,
 amsmath,amssymb,
 aps,
]{revtex4}

\usepackage{graphicx}
\usepackage{dcolumn}
\usepackage{bm}

\usepackage{graphicx}
\usepackage{slashed}
\usepackage{color}
\usepackage{amssymb}

\begin{document}

\title{Lepton mass effects in the Bethe-Heitler process}

\author{M.B.~Barbaro}
\affiliation{Dipartimento di Fisica \\ 
Universit\'a di Torino and INFN Sezione di Torino \\
Via P. Giuria 1, 10125 Torino, Italy \\
\vspace*{5pt} }
\author{C.~Maieron}
\affiliation{Laboratoire de Physique Subatomique et de Cosmologie \\
CNRS/IN2P3, Universit\'e Joseph Fourier, INPG \\
53 avenue des Martyrs, 38026 Grenoble, France}
\author{E.~Voutier}
\thanks{Corresponding author : voutier@lpsc.in2p3.fr}
\affiliation{Laboratoire de Physique Subatomique et de Cosmologie \\
CNRS/IN2P3, Universit\'e Joseph Fourier, INPG \\
53 avenue des Martyrs, 38026 Grenoble, France}

\date{\today}

\begin{abstract}
We develop the full finite lepton mass formalism for the production of real photons via the Bethe-Heitler reaction of unpolarized leptons off unpolarized nucleons. Genuine lepton mass effects are described, in particular their dependence upon the lepton mass and the initial beam energy, as well as their sensitivity to the nucleon isospin. In the minimum momentum transfer region, these effects dominate the muon induced proton cross section and become significant for electron scattering at small $x_B$.
\end{abstract}


\maketitle

\section{Introduction}
\label{Bmv-intro}

The Bethe-Heitler (BH) reaction~\cite{Bet34} is a basic process for the production of real photons. When the real photons are produced from the interaction of an incoming electron with the Coulomb field of an atom or a nucleus, it corresponds to the bremsstrahlung reaction which is commonly used to generate linearly and circularly polarized photon beams of various energies~\cite{Ols59}. When the interaction occurs with the nuclear electromagnetic field, the Bethe-Heitler process carries information about the internal structure of the target nucleus represented by its electromagnetic form factors. While the inclusive process is a radiative correction to the lepton scattering cross section~\cite{Mo69}, the exclusive process is an essential tool for the experimental determination of generalized nucleon polarizabilities~\cite{Dow11} and generalized  parton distributions (GPD)~\cite{Die03}.

In the effort to improve our understanding of the nuclear structure~\cite{Kro98} and to unravel the parton structure of nucleons and nuclei~\cite{Bel05} via the (deeply) virtual Compton scattering process ((D)VCS)~\cite{Ji97}, the BH reaction appears as a powerful contamination of the global process of the lepto-production of real photons. (D)VCS corresponds to the absorption of a virtual photon by the nucleus (parton) followed by the emission of a real photon from the excited nucleus (parton). Involving the same initial and final states the BH and (D)VCS processes interfere coherently. In the multi-GeV energy range relevant for these studies, the cross section for the lepto-production of real photons contains a strong, often dominant, contribution of the BH amplitude. The nuclear structure  information of interest is extracted~\cite{Bel02} from the unpolarized lepton cross section as a deviation from the BH contribution that should consequently be accurately subtracted. Considering the cross section difference for incident polarized leptons of opposite helicities or opposite charges, the BH process serves as a magnifier of the (D)VCS signal, the precise knowledge of its magnitude being still required to access the nuclear structure information. Specifically, the BH process is the reference reaction of the lepto-production of real photons, and any attempt to access the nuclear structure via this reaction requires a very accurate knowledge of this reference process.

In most of the available calculations of this process (see~\cite{Hau04} for a review), the mass of the incoming and scattered leptons is neglected, at variance with the crossed process (lepton pair 
production) where a few finite lepton mass calculations exist~\cite{{Tsa74},{Ber02}}. Indeed, 
the ultra-relativistic approximation seems well-justified in the energy range spanned by electron scattering experiments at MAMI~\cite{MAMI}, JLab~\cite{{Hall-A},{CLAS}} and DESY~\cite{ZEUS,H1,HERMES}, and still justified for muon scattering experiments at CERN~\cite{COMPASS}. However, a recent work on the bremsstrahlung and pair-creation processes may motivate a closer look at this approximation. Revisiting cross sections and polarization observables for these processes, this work~\cite{Kur10} showed that finite electron mass effects may result in strong differences as compared to previous massless  calculations~\cite{Ols59}. In the bremsstrahlung process, finite electron mass effects are shown to be significant in the end-point region of the cross section, and regularize polarization observables in that kinematical domain. The differences are even more striking for the pair-creation process, the reciprocal of bremsstrahlung, where the finite electron mass calculation allows for consistent polarization observables at energies as low as the reaction threshold. If it is intuitive that finite electron mass calculations should be more accurate at low incoming energy, it is less obvious to expect persistent strong differences as the energy increases. 
It is the merit of these calculations to show that additional terms involving the electron mass may significantly affect experimental observables even at relativistic energies.

The purpose of the present work is to investigate the effects of finite lepton mass on the BH process in the perspective of the on-going and future VCS and DVCS experimental programs. The next section describes the formalism for the unpolarized BH process off the nucleon and discusses the comparison with a previous massless calculation. The following sections focus on the specific features of the calculation here-below developed.

\section{Formalism}
\label{Bmv-forma}

\begin{figure}[t]
\centering
\includegraphics[bb = 125 570 300 675, scale = 1.20]{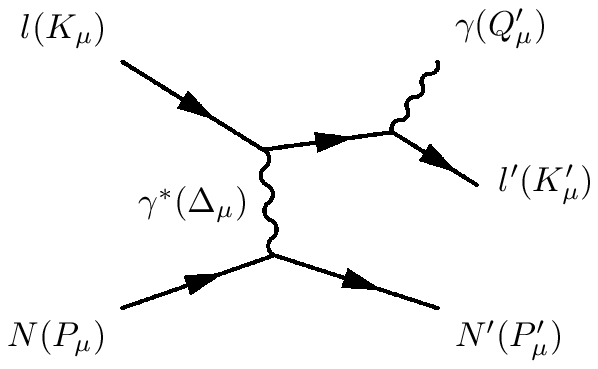}
\includegraphics[bb = 125 570 300 675, scale = 1.20]{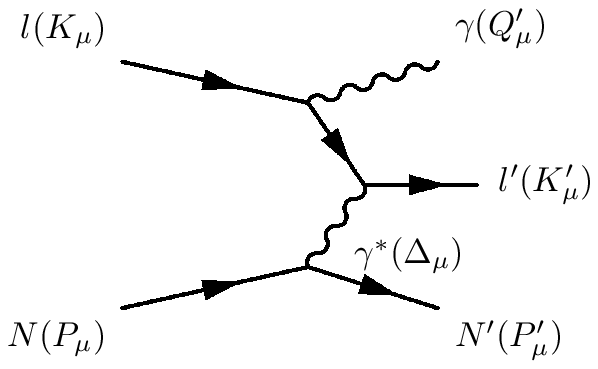}
\caption{\label{fig:FD} 
Feynmann diagrams for the Bethe-Heitler process. The four-momenta of the lepton 
$l$, nucleon $N$, virtual photon $\gamma^*$ and real photon $\gamma$ are indicated in parenthesis.}
\end{figure}

Within the Born approximation, the Bethe-Heitler elastic process
\begin{equation}
\label{eq:process}
l+N \to l'+N'+\gamma
\end{equation}
is described by the two diagrams in Fig.~\ref{fig:FD}. In the laboratory frame the four-momenta indicated in 
the figure are
$$
\begin{array}{lll}
& P_\mu = ( M, {\vec 0} )~,
& P_\mu^\prime = ( E^\prime, {\vec {p^\prime}} ) = 
\left( \sqrt{p^{\prime 2}+M^2}, {\vec {p^\prime}} \right)~, \\
& K_\mu = ( k_0, {\vec k} )= \left( \sqrt{k^2+m^2}, {\vec k} \right)~,\ \ \ 
& K_\mu^\prime = ( k_0^\prime, {\vec {k^\prime}} ) = 
\left( \sqrt{k^{\prime 2}+m^2}, {\vec {k^\prime}} \right)~, \\
& Q_\mu = K_\mu - K_\mu^\prime = ( \omega, {\vec q} )~,
& Q_\mu^\prime = ( \omega^\prime, {\vec {q^\prime}} )~, 
\end{array}
$$
where $M$ and $m$ are the nucleon and lepton mass, respectively, $Q_\mu$ is the equivalent virtual photon four-momentum corresponding to the true virtual photon of the (D)VCS process, and $Q^{\prime 2}=0$. The four-momentum transfer to the nucleon writes
\begin{equation}
\Delta_\mu  = (\Delta_0,\vec\Delta) = Q_\mu - Q_\mu^\prime = P_\mu^\prime - P_\mu \, .
\end{equation}
The corresponding elementary cross section can be expressed as~\footnote{
We use the following normalization for the lepton and hadron spinors:
$\overline{u}(p) u(p) = 2 m$ and
$\overline{U}(P) U(P) = 2 M$.
}
\begin{equation}
d^9 \sigma = \frac{1}{4kM} \frac{1}{2k^\prime_0} \frac{d^3k^\prime}{(2 \pi)^3} \frac{1}{2 \omega'} \frac{d^3q^\prime}{(2 \pi)^3} \frac{1}{2E^\prime} \frac{d^3p^\prime}{(2 \pi)^3} \, 
(2 \pi)^4 \delta^{4} ( Q_\mu - Q_\mu^\prime + P_\mu - P_\mu^\prime ) \, \left| {\mathcal T} \right|^2~.
\end{equation}
Integrating over the final nucleon and the real photon energy, the differential cross section  with respect to the final lepton three-momentum $\vec k^\prime$ and to the outgoing photon solid angle $\Omega_\gamma$ reads
\begin{equation}
\label{eq:d5s}
\frac{d^5\sigma}{d^3k^\prime d\Omega_\gamma} = \frac{1}{(2\pi)^5} \frac{1}{32 k k_0^\prime}  
\frac{1}{2M^3} \frac{{\left( 2M\omega+\Delta^2 \right)}^2} {2M\omega-\left|Q^2\right|} 
\,\, \left| {\mathcal T} \right|^2 ~.
\end{equation}
The ${\mathcal T}$-matrix for the Bethe-Heitler process is expressed in terms of the leptonic and hadronic tensors according to 
\begin{equation}
\left| {\mathcal T} \right|^2 = \frac{e^6}{\Delta^4} L_{\mu\nu} W^{\mu\nu}
\end{equation}
where
\begin{equation}
\label{eq:LW}
L_{\mu\nu} = \sum_{\lambda'} \epsilon_{\lambda'}^{\alpha}  \epsilon_{\lambda'}^{\beta *}
{\overline{\sum}}_{spin} l_{\alpha,\mu}^* l_{\beta,\nu}~,\ \ \ \ \ \ \ \ 
W^{\mu\nu} = {\overline{\sum}}_{spin}  J^{\mu *} J^\nu~.
\end{equation}
The sum over spin states stands for unpolarized electron scattering off unpolarized nucleon target. The electromagnetic nucleon current matrix element is
\begin{equation}
\label{eq:Jmu}
J^\mu = \overline{U}(P^\prime) \left[ F_1(\Delta^2) \gamma^\mu + i \frac{F_2(\Delta^2)}{2M} \sigma^{\mu\nu} \Delta_\nu 
\right] U(P)~,
\end{equation}
where $F_1$ and $F_2$ are the Dirac and Pauli electromagnetic form factors of the nucleon. $\epsilon_{\lambda'}^{\alpha}$ is the real photon polarization vector and
\begin{equation}
\label{eq:lamu}
l_{\alpha,\mu} = {\overline u(K^\prime)} \left[ 
\gamma_\alpha \Pi(K'+Q') \gamma_\mu + \gamma_\mu \Pi(K-Q') \gamma_\alpha
\right] u(K) 
\end{equation}
is the lepton current written in terms of the lepton propagator
\begin{equation}
\label{eq:Pi}
\Pi(K) = \frac{1}{{\slashed{K}} - m + i \epsilon} = \frac{\slashed{K} + m}{K^2-m^2+ i \epsilon}~.
\end{equation}
The sum over the photon polarization components reads
\begin{equation} 
\sum_{\lambda'} \epsilon_{\lambda'}^{\alpha}  \epsilon_{\lambda'}^{\beta *} = - g^{\alpha \beta} - 
\frac{Q^{\prime \alpha} Q^{\prime \beta}}{q^{\prime 2}}
\end{equation}
where the gauge term does not contribute to the scattering amplitude because of current conservation ($Q^{\prime \alpha} l_{\alpha,\mu} = 0$). Introducing 
\begin{eqnarray}
\Delta^2 P_1 = \Delta^2  - 2 K \cdot \Delta = 2 K^{\prime} \cdot Q^{\prime}  
\label{eq:prop1} \\
\Delta^2 P_2 = \Delta^2  + 2 K^{\prime} \cdot \Delta = -2 K \cdot Q^{\prime} 
\label{eq:prop2}
\end{eqnarray}
and performing the spin traces, the leptonic tensor can be recast in the following form 
\begin{eqnarray}
L_{\mu\nu} &=& \frac{8}{P_1 P_2} \, l_{\mu\nu} \label{eq:Lmunu} \\
l_{\mu\nu} &=& A\, g_{\mu\nu} + B\,\frac{K^\prime_\mu K^\prime_\nu}{\Delta^2} + C\, 
\frac{K_\mu K_\nu}{\Delta^2}  + D\, \frac{ K^\prime_\mu \Delta_\nu + \Delta_\mu 
K^\prime_\nu }{\Delta^2} + E\, \frac{ K_\mu \Delta_\nu + \Delta_\mu K_\nu }{\Delta^2} 
\nonumber \\
&+& F\, \frac{K^\prime_\mu K_\nu + K_\mu K^\prime_\nu}{\Delta^2} 
+ G\, \frac{\Delta_\mu \Delta_\nu}{\Delta^2} 
\label{eq:lmunu}
\end{eqnarray}
where
$$
\begin{array}{lll}
& A = K^{\prime 2}_\Delta + K_\Delta^2 - 2 \mu^2 \, \frac{\left[1+K^\prime_\Delta -K_\Delta \right]^2}{P_1 P_2}~,
& B = 1+\mu^2 \, \frac{P_1}{P_2}~, \\
& C = 1+\mu^2 \, \frac{P_2}{P_1}~,\ \ \ 
& D = -K^{\prime}_\Delta + 2\mu^2 \, \frac{K^{\prime}_\Delta + K_\Delta}{P_2}~, \\
& E = -K_\Delta+2\mu^2 \, \frac{K^{\prime}_\Delta + K_\Delta}{P_1}~,
& F = -G = -2\mu^2~, 
\end{array}
$$
being $\mu^2=m^2/\Delta^2$, $K_\Delta = K\cdot\Delta / \Delta^2$, and $K^\prime_\Delta = K^\prime\cdot\Delta / \Delta^2$.
The electromagnetic hadronic tensor can be written as~\cite{Raskin:1988kc}~\footnote{We only consider the symmetric part of the tensor, since the 
antisymmetric part does not contribute to the cross section for unpolarized electrons.}
\begin{equation}
W^{\mu\nu} = 4 M^2 w^{\mu\nu} = 
- 4 M^2 \, W_1 \left(g^{\mu\nu}-\frac{\Delta^\mu\Delta^\nu}{\Delta^2} \right) 
+ 4 M^2 \, W_2 V^\mu V^\nu~,
\label{eq:Wmunu}
\end{equation}
where 
\begin{equation}
V^\mu = \frac{1}{M} \left( P^\mu - 
\frac{P\cdot\Delta}{\Delta^2} \Delta^\mu \right) = 
\frac{1}{M} \left( P^\mu + \frac{\Delta^\mu}{2} \right)
\end{equation}
and 
\begin{eqnarray}
W_1 &=& \tau ( F_1 + F_2 )^2 = \tau G_M^2 
\label{eq:W1}
\\ 
W_2 &=& F_1^2 + \tau F_2^2 = \frac{1}{1 + \tau} \left( G_E^2 + \tau G_M^2 \right)\label{eq:W2}~,
\end{eqnarray}
having introduced $\tau = -\Delta^2/(4 M^2)$ and the Sachs electric and magnetic nucleon form factors~\cite{Ern60}. By contracting the reduced leptonic tensor (Eq.~\ref{eq:lmunu}) and the hadronic tensor (Eq.~\ref{eq:Wmunu}) we get
\begin{eqnarray}
l_{\mu\nu} w^{\mu\nu} & = & W_2 \left( A V^2 - B\, K^{\prime 2}_V - C\, K_V^2 - 2 F\, K_V K^\prime_V \right) 
\label{eq:lWm} \\
&-& W_1 \left[ { 3 A + B \left( \mu^2-K^{\prime 2}_\Delta\right) + C \left(\mu^2-K_\Delta^2\right)
+ F \left( 1 + 2\, K^{\prime}_\Delta - 2\,K_\Delta - 2\, K^{\prime}_\Delta K_\Delta + 2 \mu^2 \right) } \right] \nonumber
\end{eqnarray}
being $K_V = K \cdot V / \sqrt{-\Delta^2}$ and $K^\prime_V = K^\prime \cdot V / \sqrt{-\Delta^2}$. Note that $D$, $E$ and $G$ do not contribute. In the $m=0$ limit, $A = K^2_\Delta +  K^{\prime 2}_\Delta $, $B=C=1$, $F=0$, and the above expression simplifies to
\begin{equation}
\label{eq:nomass}
\left. l_{\mu\nu} w^{\mu\nu} \right|_{m=0} = \left( V^2 \, W_2 - 2 W_1 \right) \left[ K^{\prime 2}_\Delta + K^2_\Delta \right] -  
W_2 \, \left[ K^{\prime 2}_V + K^2_V  \right]~.
\end{equation}
Different contributions can be distinguished in the tensor contraction of Eq.~\ref{eq:lWm}, following the  relation
\begin{equation}
l_{\mu\nu} w^{\mu\nu} = {\cal F} = {\cal F}_0 + \mu^2 {\cal F}_2 + \mu^4 {\cal F}_4
\end{equation}
expressing a polynomial dependence in the squared reduced lepton mass $\mu$, in addition to the intrinsic mass dependence built into kinematics. The polynomial coefficients write     
\begin{eqnarray}
{\cal F}_0 & = & \alpha_1 W_1 + \alpha_2 W_2 \label{eq:F0} \\
{\cal F}_2 & = & \beta_1 W_1 + \beta_2 W_2 \label{eq:F1} \\
{\cal F}_4 & = & \gamma_1 W_1 \label{eq:F2} 
\end{eqnarray}
with
\begin{eqnarray}
\alpha_1 & = & -1 + P_1 + P_2 - \frac{P_1^2+P_2^2}{2} \label{eq:al1} \\
\alpha_2 & = &  \frac{1}{2} + \frac{k_0-k^{\prime}_0}{2M} - \frac{k_0^2+k^{\prime 2}_0}{4M^2\tau} 
- \frac{k_0  P_1 - k_0^{\prime} P_2}{2M} - \frac{P_1+P_2}{2} + \frac{P_1^2+P_2^2}{4} \, \, \, \, \, \, \label{eq:al2} \\
\beta_1 & = &  \frac{7}{4} \left( \frac{P_1}{P_2} + \frac{P_2}{P_1} \right)
+ \frac{P_1+P_2}{2} + \frac{3 P_1 P_2}{2} \label{eq:be1}\\
\beta_2 & = &  -1 - 2 \tau + \frac{k_0-k^{\prime}_0}{M} + \frac{k_0k^{\prime}_0}{M^2 \tau} 
- \frac{P_1}{P_2} \left( \frac{1}{2} + \frac{k^{\prime}_0}{2M} + \frac{k^{\prime 2}_0}{4 M^2 \tau} \right) \nonumber \\
 & & - \frac{P_2}{P_1} \left( \frac{1}{2} - \frac{k_0}{2M} + \frac{k^2_0}{4 M^2 \tau} \right) - \frac{3}{2} \left( \frac{k_0}{M} P_2 - 
 \frac{k^{\prime}_0}{M} P_1 \right) \nonumber \\
 & & - \frac{3\tau}{2} \left( P_1 P_2 - P_1 - P_2 + \frac{P_1}{2P_2} + \frac{P_2}{2P_1} \right)
\label{eq:be2} \\
\gamma_1 & = & 4 - \frac{P_1}{P_2} - \frac{P_2}{P_1} \ . \label{eq:gam1} 
\end{eqnarray}
${\cal F}_0$ is formally identical to the massless tensor contraction (Eq.~\ref{eq:nomass}) and differs from it only by the finite lepton mass kinematics. ${\cal F}_2$ and ${\cal F}_4$ are additional contributions to the cross section attached to the second and fourth power of the reduced lepton mass, respectively, and originating from the finite mass leptonic tensor (Eq.~\ref{eq:lmunu}). 
Finally, the Bethe-Heitler differential cross section with respect to the scattered lepton momentum and the real photon solid angle, for unpolarized incoming lepton and initial target, writes
\begin{equation}
\label{eq:d5sf}
\frac{d^5\sigma}{dk^\prime d\Omega_e d\Omega_\gamma} = \frac{\alpha^3}{\pi^2} \frac{1}{k M} \frac{k^{\prime 2}}{k_0^\prime}  
\frac{{\left( 2M\omega+\Delta^2 \right)}^2} {2M\omega-\left|Q^2\right|} \frac{1}{\Delta^4} \frac{1}{P_1 P_2} \,\, 
l_{\mu\nu} w^{\mu\nu}~,
\end{equation}
where $\alpha$ is the fine structure constant.

\begin{figure}[ht]
\centering
\includegraphics[scale=0.55]{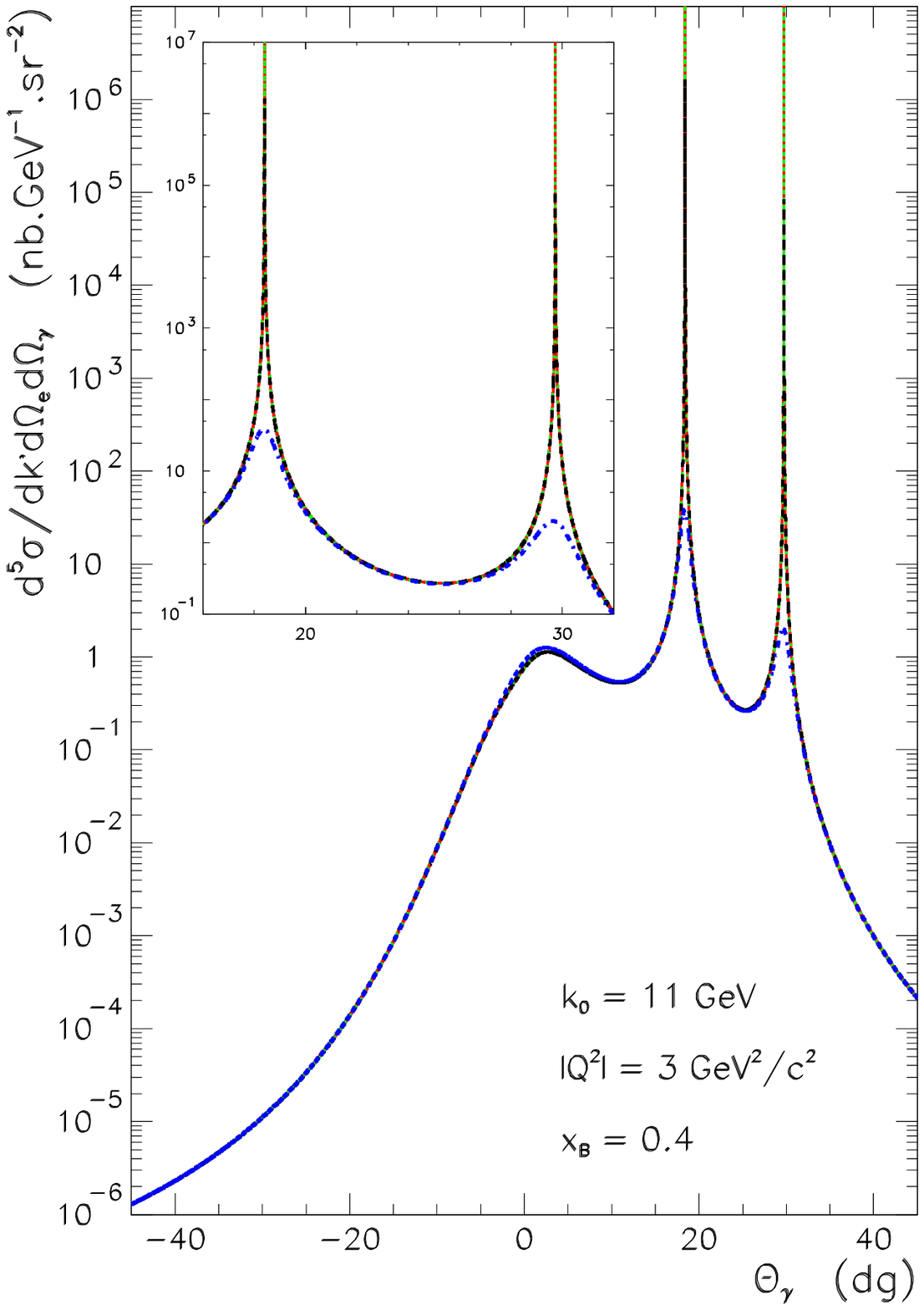}
\includegraphics[scale=0.55]{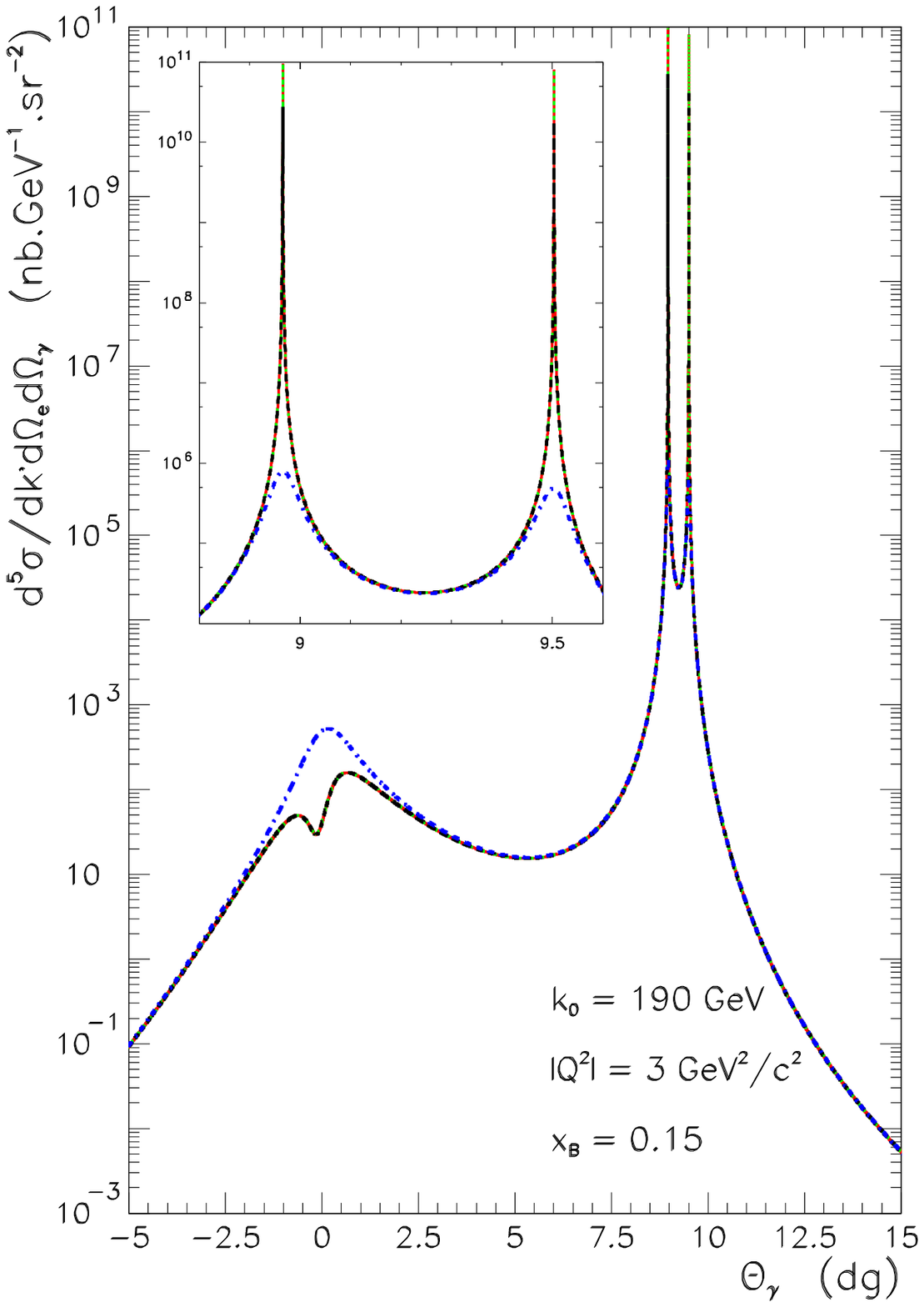}
\caption{\label{fg:bh-mass-effect} (Color online) BH differential cross section off proton at kinematics relevant of the JLab and CERN experimental program ($k_0$ is the incident lepton energy and $x_B=|Q^2|/(2M\omega)$ the Bjorken variable). The full (green) line corresponds to the massless lepton calculation of Ref.~\cite{Bel02}. The numerical results of the present work are shown for different lepton mass: dotted (red) line for massless leptons, dashed 
(black) line for electrons, and dash-dotted (blue) line for muons.} 
\end{figure}
Fig.~\ref{fg:bh-mass-effect} shows the results of the present calculations for typical JLab and COMPASS  kinematics, using the Galster parametrization for the proton electromagnetic form factors~\cite{Gal71}. The particle angles and momenta are referenced in the laboratory frame of the equivalent virtual photon ($Q$) with $\vec z$-axis parallel to $\vec q$; in that frame $\phi_{\gamma}$ is the out-of plane angle between the leptonic plane $({\vec k},{\vec {k^\prime}})$ and the hadronic plane $({\vec {q^\prime}},{\vec {p^\prime}})$. The results are displayed as function of the polar angle $\theta_{\gamma}$ of real photons for in-plane kinematics ($\phi_{\gamma}=0,\pi$). The massless lepton result of Eq.~\ref{eq:nomass} (dotted line) is first compared with the previous massless calculation of Ref.~\cite{Bel02} (full line): both approaches are in this case strictly identical, yielding the same numerical values. It is then compared with the finite mass calculations for electron- (dashed line) and muon- (dash-dotted line) induced BH. A first region is identified and singled-out in the upper left corner of the graphs, corresponding to the quasi-singularities of the intermediate leptons propagators where strong differences between massless and finite mass results are observed, particularly for muons. On the right panel, a second region is identified at lower $x_B$, corresponding to the kinematical threshold (minimum absolute momentum transfer) where significant finite mass effects are also observed. Note that the existence of these regions is in agreement with earlier observations of L.~Mo and Y.~Tsai who used similar arguments to derive specific peaking approximations for the calculation of radiative corrections to elastic and inelastic lepton scattering~\cite{Mo69}. The next sections detail the mechanisms at play in these two specific regions.

\section{Lepton propagator quasi-singularities}
\label{Bmv-lepro}

The intermediate lepton propagators entering the BH cross section are responsible for the local singular behaviour, observed on Fig.~\ref{fg:bh-mass-effect}, through the denominators ($P_1$, $P_2$). Simple algebra shows that the condition $P_1=0$ is equivalent to $K^{\prime} \cdot Q^{\prime}=0$ (similarly, $P_2=0$ is equivalent to $K \cdot Q^{\prime}=0$), leading to 
\begin{equation}
K^{\prime} \cdot Q^{\prime} = 0 \longrightarrow \cos \theta_{k^{\prime}q^{\prime}} =
\frac{\sqrt{k^{\prime 2}+m^2}}{k^{\prime}} \ .
\end{equation}
This condition is obviously never satisfied for finite mass leptons and is verified only by  massless leptons emitting a real photon in the forward direction. At fixed lepton kinematics, the $\Delta^2$-location of these in-plane quasi-singularities are 
\begin{equation}
\Delta_1^2 = - 2 M \vert Q^2 \vert \, \frac{k_0}{2 M (k_0-\omega) + \vert Q^2 \vert}
\end{equation}
for $P_1$, and
\begin{equation}
\Delta_2^2 = - 2 M \vert Q^2 \vert \, \frac{k_0 - \omega}{2 M k_0 - \vert Q^2 \vert}
\end{equation}
for $P_2$. At these momentum transfers, massless cross sections are infinite while finite mass cross sections yield finite values whose magnitudes depend on the lepton mass. This singular behaviour of the cross section was already noticed in Ref.~\cite{Bel02}. Considering finite lepton mass in the kinematics regularizes the cross section but does not allow for a precise determination of the cross section in the quasi-singularity regions, since it ignores the extra terms associated to the full leptonic tensor. 

\begin{figure}[t]
\centering
\includegraphics[scale=0.55]{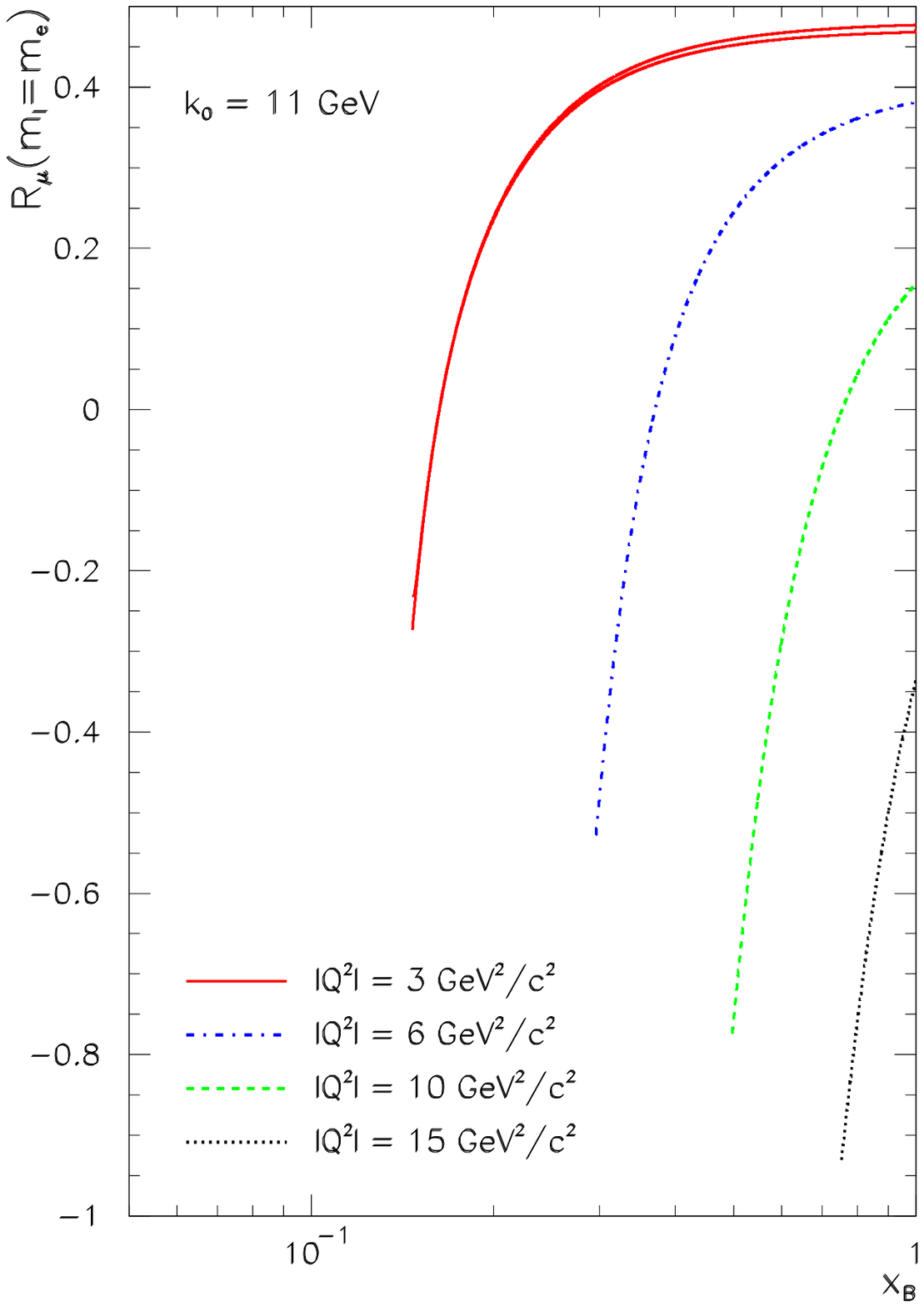}
\includegraphics[scale=0.55]{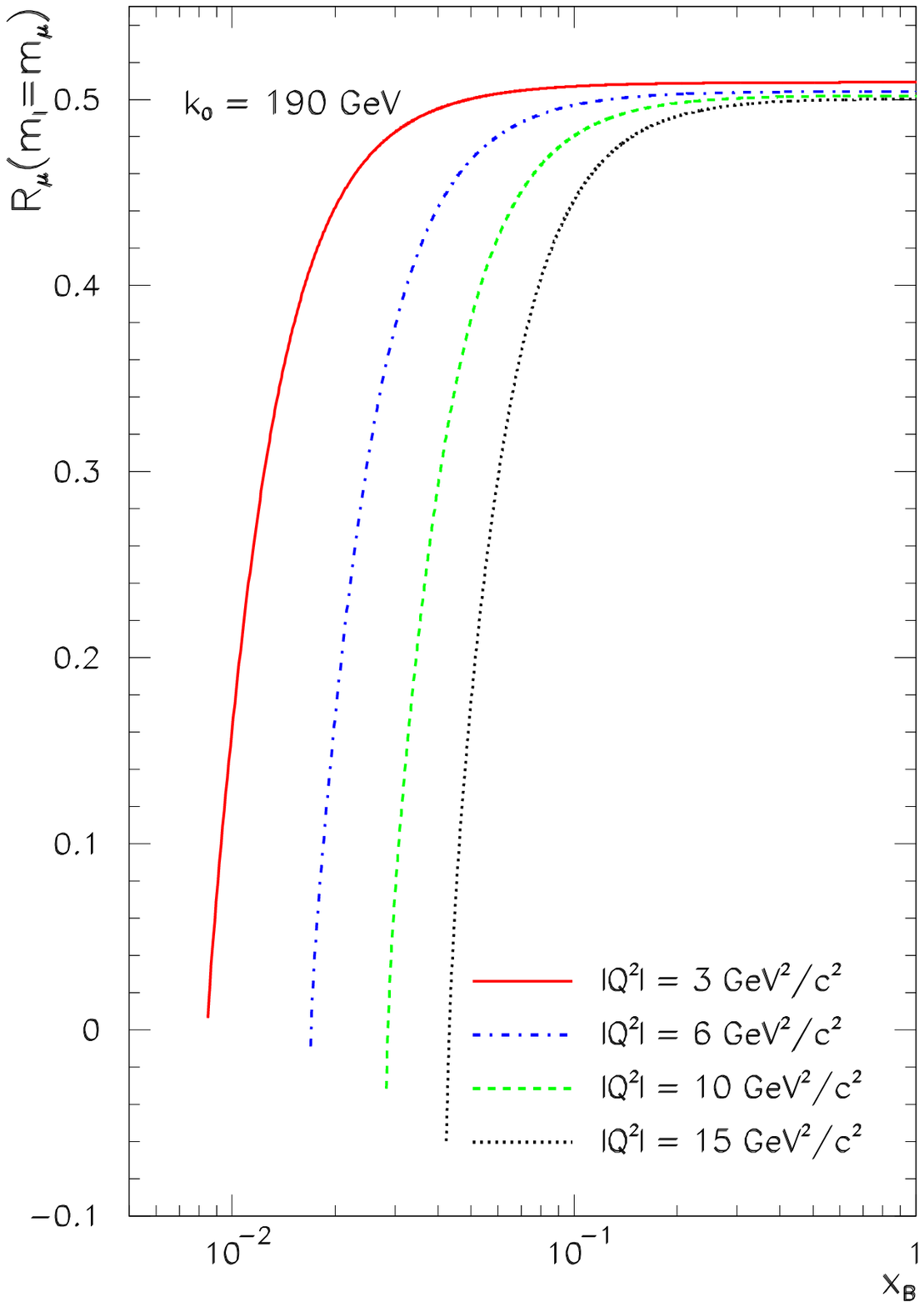}
\caption{(Color online) $x_B$ evolution of the ratio $R_{\mu}$ for electron (left) and muon (right)  scattering off proton calculated at the scattered lepton  
quasi-singualirity $\Delta^2_1$ for different initial energy and momentum transfer of the equivalent virtual photon. The additional full line on the left panel (upper curve) indicates the muon ratio at $\vert Q^2 \vert=3$~GeV$^2$/$c^2$; other $\vert Q^2 \vert$ cannot be distinguished 
from the electron ratio.} 
\label{bh-pole}
\end{figure}
The importance of extra contributions to the leptonic tensor specific of the finite lepton mass formalism can be expressed by the ratio
\begin{equation}
R_{\mu} = \frac{ {\cal F} - {\cal F}_0 }{{\cal F}_0}~,
\label{eq:ratmu}
\end{equation}
identical to the corresponding differential cross section ratio. Fig.~\ref{bh-pole} shows the $x_B$ evolution of this ratio calculated for typical JLab and COMPASS lepton kinematics at the quasi-singularity $\Delta^2_1$ corresponding to emission of real photons in the direction of scattered leptons. Globally, the extra terms ${\cal F}_2$ and ${\cal F}_4$ are significantly contributing to the cross section at the quasi-singularity. Notably, $R_{\mu}$ exhibits a saturation behaviour more or less rapidly reached depending on kinematics, the saturation value being roughly independent of the lepton mass. Eq.~\ref{eq:F0}-\ref{eq:gam1} help understand these features. At the quasi-singularity $\Delta^2_1$, $P_1$ is quasi-null and extra contributions can be reduced to their dominant contribution in $1/P_1$
\begin{eqnarray}
{\cal F}_2 & \simeq  & \frac{P_2}{P_1} \, \left[ \frac{7}{4} \, W_1 - \left(  \frac{1}{2} + \frac{3 \tau}{4} - 
\frac{k_0}{2M} + \frac{k^2_0}{4 \tau M^2} \right) W_2 \right]  \\
{\cal F}_4 & \simeq  &  - \frac{P_2}{P_1} \, W_1 \ .
\end{eqnarray}
Moreover, from Eq.~\ref{eq:prop1} and Eq.~\ref{eq:prop2} one obtains at $\Delta^2_1$ 
\begin{eqnarray}
P_1 & \simeq  &  \frac{q^{\prime}}{k^{\prime}} \, \mu^2 \\
P_2 & \simeq  &  \frac{q^{\prime}}{k^{\prime} + q^{\prime}} \, ,
\end{eqnarray}
such that the $\mu^2$ dependence of $P_2 / P_1$ leads to a lepton mass independent ratio $R_{\mu}$ 
up to $\mu^4$ order. Assuming further $x_B=1$, leading to $ q^{\prime}=0$ and  $\Delta^2 = Q^2$, we obtain
\begin{eqnarray}
\mu^2 {\cal F}_2 + \mu^4 {\cal F}_4 & \simeq  & \frac{7}{4} \, W_1 - \left(  \frac{1}{2} + \frac{3 \tau}{4} - \frac{k}{2M} + \frac{k^2}{4 \tau M^2} \right) W_2 \\
{\cal F}_0 & \simeq  & - W_1 + \left( \frac{1}{2} + \frac{k}{M} - \frac{k^2}{2 \tau M^2} \right) W_2 
 \ . 
\end{eqnarray}
In the approximation of the dominance of the magnetic form factor contribution to the cross section, the electric contribution to $W_2$ in Eq.~\ref{eq:W2} can be neglected and Eq.~\ref{eq:ratmu} becomes
\begin{equation}
R_{\mu} \simeq  \frac{1}{2} - \frac{3}{1 + \frac{\tau}{\tau + 1} {\left( \frac{k}{M \tau} - 1 \right)}^2}  \ .
\label{eq:Rmu1}
\end{equation}
This expression shows that $R_{\mu}$ reaches a maximum value about $1/2$ at small $\tau$ and reproduces the $Q^2$-dependence observed on the left panel of Fig.~\ref{bh-pole} at $x_B=1$. A minimum value about $-1$ is also reproduced at the maximum kinematically allowed $Q^2$, consistently with exact calculations. Similar arguments can also be developed for the quasi-singularity at $\Delta^2_2$ corresponding to emission of real photons in the lepton beam direction. 

To conclude this section, it is worth stressing that an accurate evaluation of the BH cross section in the quasi-singularity regions requires the higher order terms in $\mu^2$ specific of the finite lepton mass formalism, independently of the kinematics of the reaction and most notably even in the ultra-relativistic case.

\section{Kinematical threshold region}
\label{Bmv-kinem}

\begin{figure}[t]
\centering
\includegraphics[scale=0.55]{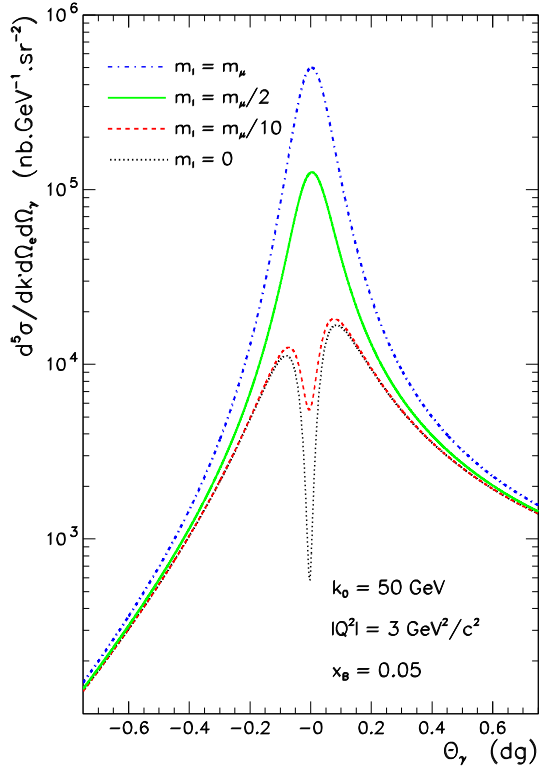}
\caption{(Color online) BH differential cross section off proton in the $\Delta^2_{min}$ region at fixed kinematics and for different incident lepton masses.} 
\label{bh-ml}
\end{figure}

Significant finite lepton mass effects are also identified at minimum momentum transfer $\Delta^2_{min}$ (right panel of Fig.~\ref{fg:bh-mass-effect}), a region of particular interest for 
the experimental determination of the Ji sum rule~\cite{Ji97} leading to the measurement of parton contribution to the total angular momentum of the nucleon. At this momentum transfer, real photons are  emitted in the direction of the equivalent virtual photon with an energy
\begin{equation}
\omega^{\prime} = \omega + \frac{\Delta^2_{min}}{2M} 
\end{equation}
where  
\begin{equation}
\Delta^2_{min} = - M \frac{ {\left( \omega - q \right)}^2}{\omega - q + M} \, .
\end{equation}
Fig.~\ref{bh-ml} shows that the BH differential cross section in the $\Delta^2_{min}$ region is strongly sensitive to the lepton mass, several orders of magnitude distinguishing massless and finite mass calculations. These effects originate again from the extra terms of the leptonic tensor.  

\begin{figure}[t]
\centering
\includegraphics[scale=0.55]{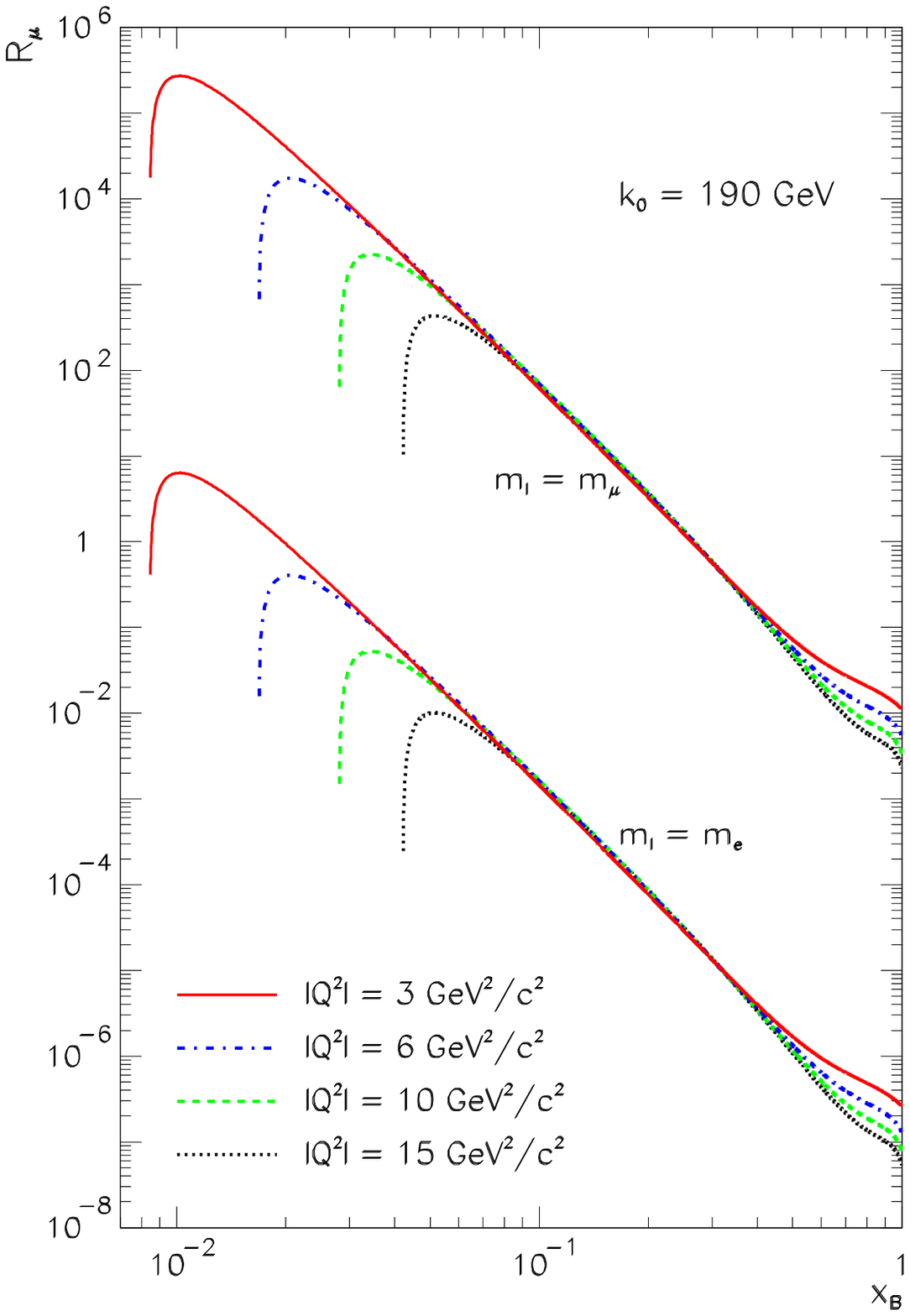}
\includegraphics[scale=0.55]{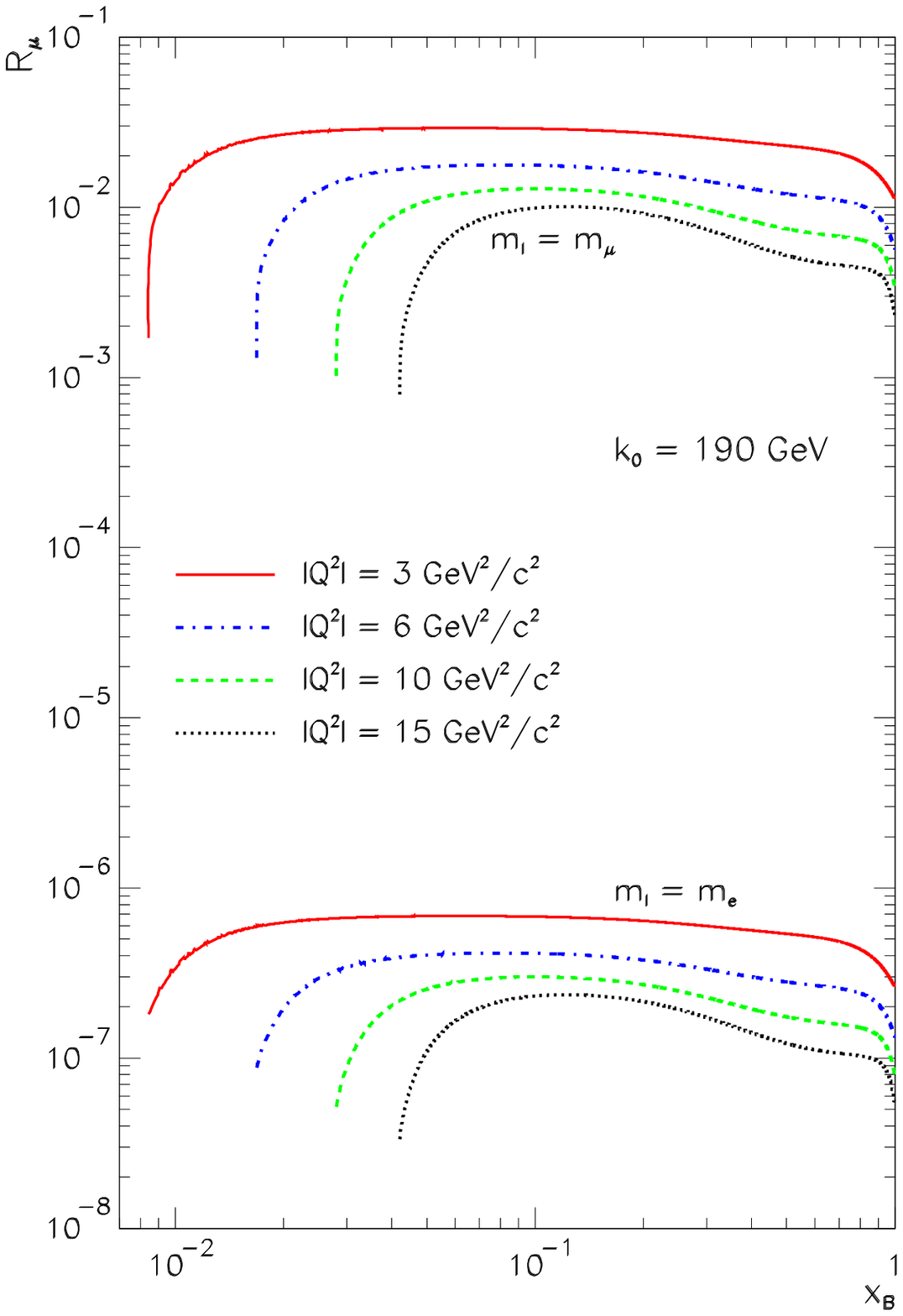}
\caption{(Color online) $x_B$ evolution of the ratio $R_{\mu}$ calculated for different $Q^2$ at $\Delta^2_{min}$ and an incident beam energy of 190~GeV of muons (upper curves) and electrons (lower curves): proton (left panel), neutron (right panel).} 
\label{bh-mass-tmi}
\end{figure}

The ratio $R_{\mu}$ (Eq.~\ref{eq:ratmu}) characterizing the deviation of the finite mass with respect to the massless leptonic tensor is shown on Fig.~\ref{bh-mass-tmi} for different equivalent virtual photon squared four-momentum at the kinematical point $\Delta^2_{min}$, for an incident beam energy of 190~GeV. The muon ratio for a proton target (upper curves of the left panel) shows very large effects in the small $x_B$ region. The electron ratio (lower curves of the left panel) exhibits the same behaviour and scales with the muon ratio as the squared lepton mass ratio. Most notably, $R_{\mu}$ appears almost independent of $Q^2$ except in the region of the kinematical boundaries of the phase space. Additionally, the initial beam energy has little influence on this behaviour and  essentially acts as a phase space magnifier: higher beam energies open access to smaller $x_B$ where larger ratios are obtained. It is interesting to note that extra contributions effects appear negligible for a neutron target (right panel of Fig.~\ref{bh-mass-tmi}).

These features can be understood by considering a simplified expression of the leptonic tensor. Introducing the physics scale parameter $z = M^2 x_B^2 / \vert Q^2 \vert$, the dimensionless lepton energy $\zeta = k_0 / M$, and $\tau_{min} = - \Delta^2_{min}/(4M^2)$, we have 
\begin{eqnarray}
4 \tau_{min} & = & \frac{x_B^2}{x_B\,(1-x_B) + z} \, \left[ 1 + \frac{(1-x_B) (1-\sqrt{1+4 z})}{2 z} \right] \\
P_1 & = &  - \, ( 1 - x_B) \, \left[ \frac{1}{\sqrt{1+4z}} + \frac{\zeta}{x_B} \left( 1 + \frac{1}{\sqrt{1+4z}} \right) \right] \\
P_2 & = &  - \, ( 1 - x_B) \, \left[ \frac{1}{\sqrt{1+4z}} + \left( \frac{1}{2z} - \frac{\zeta}{x_B} \right) 
\left( 1 + \frac{1}{\sqrt{1+4z}} \right) \right] \, .
\end{eqnarray}
In the limit $k_0 \gg \omega$, equivalent to $\zeta/x_B \gg 1/2z$, the lepton propagators reduce to  
\begin{equation}
P_1  \simeq - P_2 \simeq 2 \, \frac{x_B-1}{x_B} \, \zeta \, .
\end{equation}
Replacing these expressions in Eqs.~\ref{eq:al1}-\ref{eq:gam1}, one obtains at leading order in 
$\zeta$   
\begin{eqnarray}
\alpha_1 & \simeq & - 4 \, {\left( \frac{x_B-1}{x_B} \right)}^2 \, \zeta^2 \\
\alpha_2 & \simeq &   2 \, \left[ {\left( \frac{x_B-1}{x_B} \right)}^2 - \frac{x_B-1}{x_B} - \frac{1}{4\tau_{min}} \right] \, \zeta^2 \\
\beta_1  & \simeq & -6 \, {\left( \frac{x_B-1}{x_B} \right)}^2 \, \zeta^2 \\
\beta_2 & \simeq & 6 \, \left[  \frac{x_B-1}{x_B} + \tau_{min} {\left( \frac{x_B-1}{x_B} \right)}^2 + \frac{1}{4\tau_{min}} \right] \, \zeta^2 \\
\gamma_1 & \simeq & 6 \, .
\end{eqnarray}
It is readily seen from these relations that the leptonic tensor ratio (Eq.~\ref{eq:ratmu}) is approximately independent of the initial beam energy and of $Q^2$-dependence. The additional approximation $z \ll x_B$, valid in the kinematical region $Q^2 \gg M^2$, leads to 
\begin{eqnarray}
\frac{1}{4 \tau_{min}} & \simeq & \frac{x_B^2}{1-x_B} \\
\alpha_2 & \simeq &  0 \\
\beta_2 & \simeq & 6 \, {\left( \frac{x_B-1}{x_B} \right)}^2 {\left( 1 + \tau_{min} \right)} \, \zeta^2 \, . 
\end{eqnarray}
The leptonic tensor ratio then writes
\begin{equation}
R_{\mu} \simeq \frac{3}{2} \, \frac{m^2}{M^2} \, \frac{1-x_B}{x_B^2} \, \frac{W_1-{\left[ 1 + \tau_{min} \right]} W_2}{W_1} \simeq  6 \, \frac{m^2}{M^2} \, \frac{(1-x_B)^2}{x_B^4} \, \frac{G_E^2}{G_M^2} \, .
\label{eq:ratfin}
\end{equation}
This approximate expression shows the strong $x_B$-dependence of the ratio responsible for the large effects observed at small $x_B$, and provides an explanation for its sensivity to the nucleon isospin (Fig.~\ref{bh-mass-tmi}). It also shows that mass effects become sizeable ($R_{\mu} > 1$) for the 
proton when $x_B<0.26$ for muon scattering and $x_B<0.02$ for electron scattering, in agreement with the exact results of Fig.~\ref{bh-mass-tmi}. It may be somehow surprising that finite lepton mass effects  appear sensitive to the nucleon target isospin. It is indeed quite unconventional that finite mass effects are, as in the present case, linked to the electromagnetic structure of the nucleon. The difference between the BH cross section for 
protons and neutrons expresses not only in the magnitude of the massless lepton cross section but also
in the sensitivity to mass corrective terms, as seen from the ratio dependence on the electric form factor (Eq.~\ref{eq:ratfin}) which vanishes at small momentum transfer for the neutron. 

Summarizing, extra contributions to the leptonic tensor have been shown to dominate the BH differential cross section off protons in the minimum momentum transfer region. This effect is definitely important for muon scattering and becomes significant for electron scattering when the beam energy is high enough to allow exploring the region of $x_B$ below a few $10^{-2}$, that is for example, the kinematical range of the next generation of electron-ion colliders~\cite{Acc12}.

\section{Conclusions}
\label{Bmv-concl}

This work presents a full calculation of the BH process of unpolarized leptons off unpolarized nucleons where, at variance with previous calculations, the initial and final lepton masses are not neglected. The inclusion of the finite lepton mass has been shown to generate additional contributions to the cross section that become determinant in two specific domains: 
\begin{itemize}
\item{Lepton propagator quasi-singularities: here extra contributions to the 
leptonic tensor are required for an accurate determination of the BH cross section, independently of the lepton considered;}
\item{Minimum momentum transfer regions: in the $\Delta^2_{min}$ region, of particular interest for the experimental determination of the Ji sum rule, extra contributions dominate muon scattering off protons and must also be considered for electron scattering when $x_B$ becomes significantly small (a few $10^{-2}$). Moreover, they cancel for a neutron target as a consequence of their dependence on the electric nucleon form factor.}
\end{itemize}
The BH cross section is generally sensitive to the parametrization of the nucleon electromagnetic form factors, however the magnitude of lepton mass effects as expressed by the leptonic tensor ratio (Eq.~\ref{eq:ratmu}) is only weakly depending on the model. This dependence increases with the four-momentum transfer in the quasi-singularity region, while it does not show up in the minimum momentum region because of the smallness of the four-momentum transfer. Before concluding it is worth emphasizing that the effects investigated in this work are genuine finite mass effects in the sense that they do not rely on the extreme relativistic limit of the kinematics, usually employed in high energy calculations, but are related to extra terms appearing in the leptonic tensor. Moreover, their magnitude increases with the beam energy, in opposition to the ultra-relativistic limit. These effects have been shown to be very sizeable in the above mentioned  kinematic regions and may affect the quantitative analysis of high energy lepton scattering experimental DVCS data.

\section{Acknowledgments}
\label{Ackn}

We would like to thank Markus Diehl for stimulating discussions. This work was supported by the Italian Istituto Nazionale di Fisica Nucleare under contract MB31, by the French Centre National de la Recherche Scientifique and by the GDR 3034 Chromodynamique Quantique et Physique des Hadrons.


\newpage  

\begin{center}
{\Large\bf{Lepton mass effects in the Bethe-Heitler process}} \\ {\small\bf{(Errata to Phys. Lett. B 726 (2013) 505)}}
\end{center}

\vspace*{10pt}

{\large{M.B.~Barbaro, C.~Maieron, E.~Voutier}}

\vspace*{10pt}

\begin{figure}[h]
\centering
\includegraphics[scale=0.55]{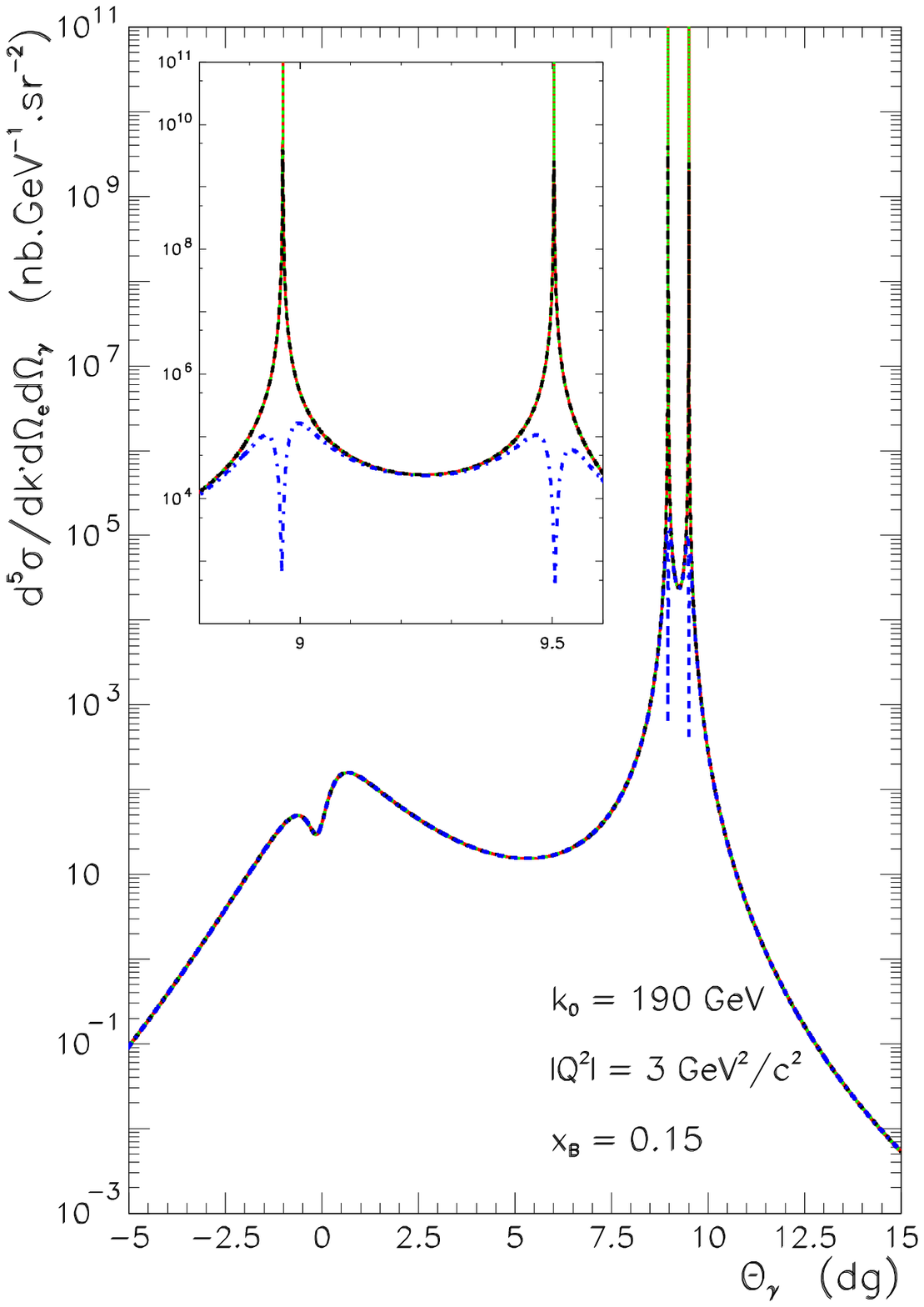}
\includegraphics[scale=0.55]{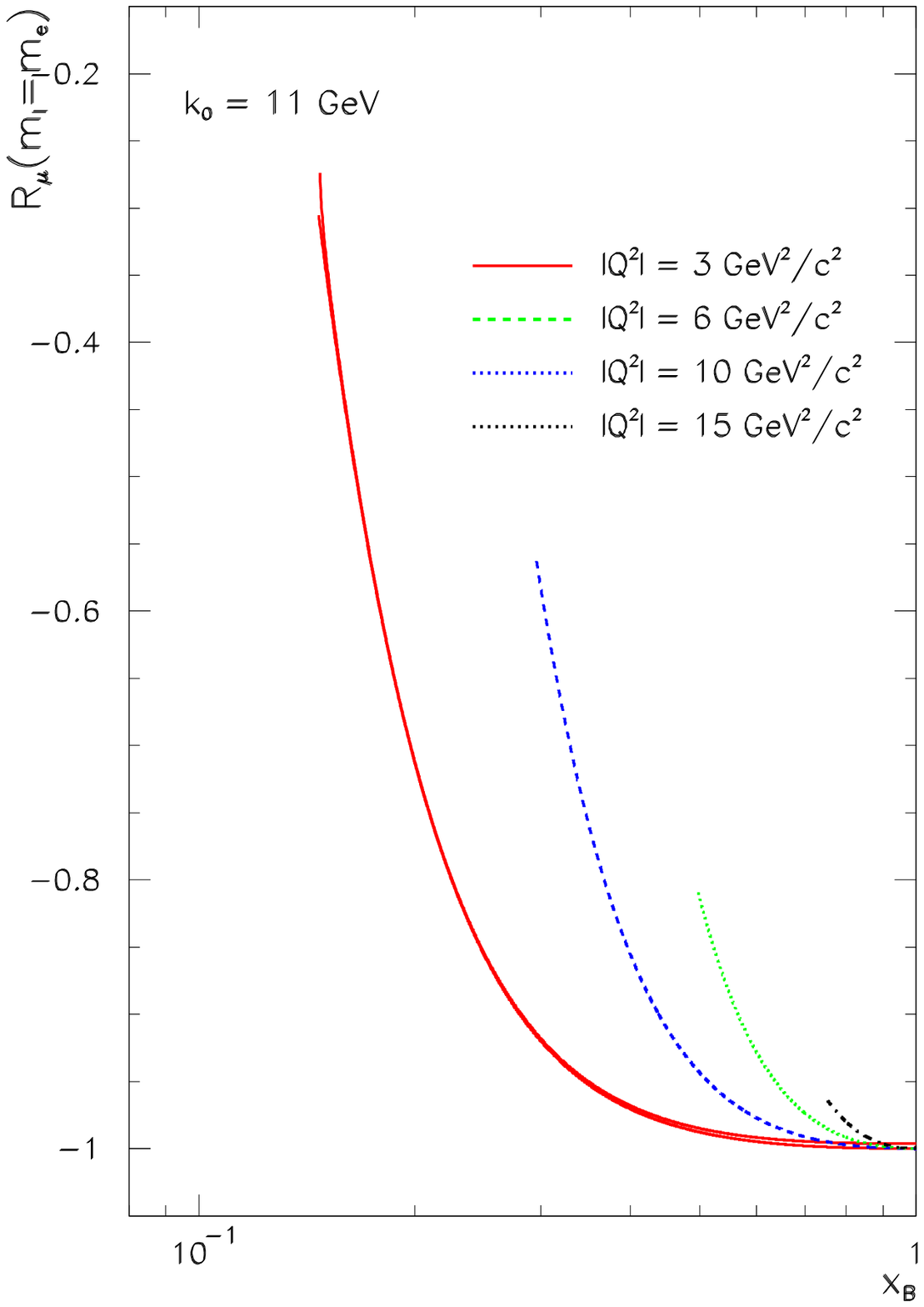}
\caption{\label{fignew} (Color online) BH differential cross section off proton in replacement of the right panel of 
Fig.~2, and $R_{\mu}$ evolution at $\Delta^2_1$ in replacement of the left panel of Fig.~3 }
\end{figure}

The correct expressions for the coefficients $B$ and $C$ in the leptonic tensor defined on page 506 are
$$ B = 1-2\mu^2 \, \frac{P_1}{P_2} \,\,\,\,\, \mathrm{and} \,\,\,\,\, C = 1-2\mu^2 \, \frac{P_2}{P_1} \, . $$
As a consequence, Eqs.~27-29 should be replaced with
\begin{eqnarray}
\beta_1 & = &  \frac{P_1}{P_2} + \frac{P_2}{P_1} + 2 \left(P_1+P_2\right) \label{eq:be1new} \nonumber \\ 
\beta_2 & = &  -1 - 2 \tau + \frac{k_0-k^{\prime}_0}{M} + \frac{k_0 k^{\prime}_0}{M^2 \tau} 
- \frac{P_1}{P_2} \left( \frac{1}{2} - \frac{k^{\prime}_0}{M} - \frac{k^{\prime 2}_0}{2 M^2 \tau} \right) 
 - \frac{P_2}{P_1} \left( \frac{1}{2} + \frac{k_0}{M} - \frac{k^2_0}{2 M^2 \tau} \right) 
\label{eq:be2new} \nonumber \\
\gamma_1 & = & 4 + 2 \left(\frac{P_1}{P_2} - \frac{P_2}{P_1}\right) \ . \nonumber  
\end{eqnarray}
The main effects of these corrections are shown on Fig.~\ref{fignew}. Our conclusions remain unchanged in the quasi-singularity regions, while the cross section enhancement observed at minimum transfer cancels. For completeness, 
Eqs.~35-36 should also be replaced with
\begin{eqnarray}
{\cal F}_2 & \sim & \frac{P_2}{P_1} \, \left[ W_1 - \left(  \frac{1}{2} + 
\frac{k_0}{M} - \frac{k^2_0}{2 M^2 \tau } \right) W_2 \right]  \nonumber \\
{\cal F}_4 & \sim &  2\,\frac{P_2}{P_1} \, W_1 \nonumber
\end{eqnarray}
leading to the substitution of Eq.~39 by
\begin{equation}
\mu^2 {\cal F}_2 + \mu^4 {\cal F}_4 \sim \ W_1 - \left(  \frac{1}{2} +  
\frac{k}{M} - \frac{k^2}{2 M^2 \tau } \right) W_2 + {\cal O}(\mu^2) \nonumber
\end{equation}
and Eq.~41 with 
$$ R_{\mu} \sim -1 \, .$$

We are deeply grateful to N.~d'Hose and P.A.M.~Guichon for making available unpublished results, and helping us to 
find the source of discrepancy between our works.

\end{document}